\documentclass[manuscript]{aastex63}
\usepackage{graphicx}
\usepackage{supertabular}
\usepackage{longtable}
\usepackage{CJK}
\usepackage{tikz}
\usepackage{threeparttable}
\usepackage{amsmath}
\usetikzlibrary{shapes,arrows}
\received{----}
\revised{----}
\accepted{Mar 12, 2025}
\submitjournal{AJ}

\begin{document}
\begin{CJK*}{UTF8}{gbsn}
\title{Extinction Distributions in Nearby Star-resolved Galaxies. I. M31}

\author[0000-0003-3860-5286]{Yuxi Wang (王钰溪)}
\affiliation{Department of Astronomy, College of Physics and Electronic Engineering, Qilu Normal University, Jinan 250200, People's Republic of China; \rm{\href{yuxiwang@qlnu.edu.cn}{yuxiwang@qlnu.edu.cn}}; \rm{\href{yiren@qlnu.edu.cn}{yiren@qlnu.edu.cn}}}

\author[0000-0003-4195-0195]{Jian Gao (高健)}
\affiliation{Institute for Frontiers in Astronomy and Astrophysics, Beijing Normal University, Beijing 102206, People's Republic of China; \rm{\href{jiangao\@bnu.edu.cn}{jiangao@bnu.edu.cn}}}
\affiliation{School of Physics and Astronomy, Beijing Normal University, Beijing 100875, People's Republic of China}

\author[0000-0003-1218-8699]{Yi Ren (任逸)}
\affiliation{Department of Astronomy, College of Physics and Electronic Engineering, Qilu Normal University, Jinan 250200, People's Republic of China; \rm{\href{yuxiwang@qlnu.edu.cn}{yuxiwang@qlnu.edu.cn}}; \rm{\href{yiren@qlnu.edu.cn}{yiren@qlnu.edu.cn}}}

\author[0000-0003-2472-4903]{Bingqiu Chen (陈丙秋)}
\affiliation{South-Western Institute for Astronomy Research, Yunnan University, Kunming, 650500, People's Republic of China; \rm{\href{bchen\@ynu.edu.cn}{bchen@ynu.edu.cn}}}



\begin{abstract}
An extinction distribution of the Andromeda Galaxy (M31) is constructed with member stars as tracers by fitting multiband photometric data from UKIRT/WFCAM, PS1, and Gaia DR3.
The resulting extinction distribution covers approximately 10 deg$^2$ of M31 with a resolution of approximately 50 arcsec, providing the largest coverage to date based on stellar observations.
The derived average extinction, $A_V = 1.17$ mag, agrees well with previous studies. 
To account for foreground extinction, an extinction map of the Milky Way toward M31 with a resolution of $\sim$ 1.7 arcmin is also constructed, yielding an average extinction of $A_V \approx 0.185$ mag. 
The results offer a valuable tool for extinction correction in future observations, such as those from the China Space Station Telescope, and provide insights for improving dust models based on the spatial distribution of dust in galaxies like M31.
\end{abstract}



\section{Introduction} \label{sec:intro}

Interstellar dust is a crucial component of the Milky Way (MW) and external galaxies, despite constituting only about 1\% of the interstellar medium's mass \citep{2003ARA&A..41..241D}. 
It plays a significant role in star formation and is released into the interstellar medium during the final stages of stellar evolution, regulating the thermodynamic properties of interstellar gas and influencing the elemental abundances of galaxies \citep{2022HabT.........1G}. 
Additionally, dust impacts the observational properties of galaxies \citep{2003ARA&A..41..241D}, making its study essential for understanding galactic evolution and structure \citep{2018ARA&A..56..673G}.

Interstellar dust was first discovered through the absorption and scattering of light \citep{1930PASP...42..214T}, a process now known as interstellar extinction. 
Extinction is a vital tool for probing dust properties such as composition and size distribution \citep{2003ARA&A..41..241D,2018ARA&A..56..673G}. 
While most of our understanding comes from the MW, studying extinction in nearby galaxies provides unique constraints on dust processes due to the varied interstellar environments and higher spatial resolution compared to distant galaxies \citep{2018ARA&A..56..673G,2022HabT.........1G}. 
Mapping dust extinction not only aids in correcting astronomical observations but also offers insights into the variation of dust properties across different interstellar and galactic environments.
Extinction maps are also vital for accurate measurements of stellar populations and distances within galaxies \citep{2021ApJS..252...23S}, and play a key role in the study of galaxy formation and star formation history \citep{2020ARA&A..58..529S}.
In addition to the scientific importance, extinction distribution has practical applications, particularly with the upcoming China Space Station Telescope (CSST) \citep{2021CSB...111.111C}, which will offer multiband imaging and slitless spectroscopy surveys with unprecedented depth, coverage, resolution, and near-ultraviolet capabilities. 
Mapping the extinction in nearby galaxies will be essential for preparing accurate corrections for CSST data and supporting a broad range of astrophysical research.

The Large Magellanic Cloud (LMC) and Small Magellanic Cloud (SMC), the MW's largest satellite galaxies, have been extensively studied for their dust extinction distributions.
These studies, using different tracers (e.g., RR Lyrae stars: \citealt{2011AJ....141..158H,2009ApJ...704.1730P}, Cepheids: \citealt{2016ApJ...832..176I,2019A&A...628A..51J}, red clump stars: \citealt{2021ApJ...910..121N,2021ApJS..252...23S}, background galaxies: \citealt{2022MNRAS.516..824B}, stars from surveys: \citealt{2022MNRAS.511.1317C}), have produced consistent extinction maps.

The Andromeda Galaxy (M31), due to the proximity, is a prime target for dust studies \citep{2015ApJ...815...14C,2022ApJS..259...12W}, as well as the other researches closely related to dust in the galaxy (e.g., \citealt{2021ApJ...907...18R,2021ApJ...912..112W}), such as stellar population, star formation history, etc. 
Recent efforts, such as those by \citet{2014ApJ...780..172D}, have mapped M31's dust mass surface density using data from the Spitzer Space Telescope and Herschel Space Observatory, from which an extinction distribution can be inferred based on a dust model \citep{2007ApJ...657..810D}.
This emission-based approach, however, has been criticized for overestimating dust mass. 
Instead, extinction mapping using near-infrared color-magnitude diagrams (CMD) of red giant branch (RGB) stars, such as the work of \citet{2015ApJ...814....3D}, offers higher resolution but covers only $\sim$ 1/3 of M31's star-forming disk.
High-resolution study has also been conducted in the bulge of M31, though this too is spatially constrained \citep{2016MNRAS.459.2262D}.
In summary, current extinction distributions of M31 are limited in terms of spatial coverage or the adopted dust model, often leading to inaccurate extinction corrections, which in turn affect studies of stellar populations and star formation history. 
Improving the extinction map of M31 is crucial for accurately recovering the intrinsic brightness of its celestial objects.

In this work, we use member stars of M31 from \citet{2021ApJ...907...18R} as tracers. 
By fitting the observed photometric data from optical to near-infrared wavelengths, we construct an updated extinction map of M31 that covers a larger sky area than previous maps. 
We also account for Galactic foreground extinction, which is mapped separately. 
Section \ref{sec:data} details the sample and data used, Section \ref{sec:method} outlines our methodology, and Section \ref{sec:re} presents our results and discussion. 
The conclusions are summarized in Section \ref{sec:conclusion}.


\section{Data and Sample} \label{sec:data}

As discussed earlier, extinction distributions in galaxies can be mapped using member stars as tracers, which provide more accurate and effective results compared to dust emission-based methods \citep{2015ApJ...814....3D}. 
The main limitation of current extinction maps in M31 is the lack of tracer samples with both low contamination and high completeness.
One challenge is the significant contamination from foreground dwarf stars, which can affect up to 30\% - 50\% of stellar samples in external galaxies \citep{2009ApJ...703..420M, 2021ApJ...907...18R}. 
This contamination introduces errors in the extinction calculations for M31. 
Additionally, the large distance to M31 results in lower tracer density, as low-luminosity stars cannot be observed. 
Higher tracer density improves the resolution of extinction distributions, enabling more accurate recovery of intrinsic brightness in M31.
Fortunately, \citet{2021ApJ...907...18R} established a reliable sample of member stars in M31 by removing foreground dwarfs in the $J-H/H-K$ diagram from data collected by the Wide Field Camera (WFCAM) on the 3.8 m United Kingdom Infra-Red Telescope (UKIRT) \citep{2013ASSP...37..229I}.

In this work, we use the 823, 189 member stars from \citet{2021ApJ...907...18R} as extinction tracers to construct an updated extinction distribution of M31. 
The sample is cross-matched with photometric data from UKIRT/WFCAM \citep{2013ASSP...37..229I}, the Panchromatic Survey Telescope and Rapid Response System release 1 (PS1) Survey \citep{2016arXiv161205560C}, and Gaia Data Release 3 (Gaia DR3; \citealt{2023A&A...674A...1G}).

The WFCAM on the UKIRT performs images in three near-IR bands: $J$ (1.25 $\mu$m), $H$ (1.63 $\mu$m), and $K$ (2.20 $\mu$m)\citep{2008A&A...487..131C,2013ASSP...37..229I,2020ApJ...889...44N}.
We obtain the $JHK$ magnitudes of the tracers from the catalog in \citet{2021ApJ...907...18R} and require that the photometric errors for these bands be less than 0.1 mag.
As a result, the photometry of 720, 906 tracers is maintained.

The Pan-STARRS is an innovative wide-field astronomical imaging and data processing facility \citep{2002SPIE.4836..154K,2010SPIE.7733E..0EK}, that surveys the sky in five broadband filters from 0.4 to 1.0 $\mu$m ($g$, $r$, $i$, $z$, $y$) \citep{2010ApJS..191..376S,2012ApJ...756..158S,2012ApJ...750...99T}.
For the PS1 data, we select high-quality photometry by filtering for reliable flags in the catalog and exclude data with magnitude errors greater than 0.1 mag.
Consequently, photometric data of PS1 is adopted for 395, 799 stars in the sample.

Gaia DR3 is the third data release of the European Space Agency's Gaia mission \citep{2016A&A...595A...1G}, and includes astrometry and broad band photometry ($G$, $G_{\rm BP}$, and $G_{\rm RP}$ ) for a total of 1.8 billion objects \citep{2023A&A...674A...1G}.
Due to the broad coverage of the $G$ band (0.33 - 1.05 $\mu$m), we exclude it from our analysis and use only the $G_{\rm BP}$ and $G_{\rm RP}$ bands, ensuring that their photometric errors are less than 0.1 mag.
Finally, Gaia DR3 photometry for only 6055 stars in the sample is retained.

In total, up to 10 photometric bands (optical to near-IR) are available for each tracer.
The observed color indexes ($g-r$, $r-i$, $i-z$, $z-y$, $y-J$, $J-H$, $H-K$, $G_{\rm BP}-G_{\rm RP}$) are used to map the dust extinction in M31.

Additionally, the foreground MW extinction cannot be neglected when calculating extinction in external galaxies. 
The extinction law in M31 differs from that of the MW \citep{1996ApJ...471..203B,2014ApJ...785..136D,2015ApJ...815...14C,2022ApJS..259...12W}. 
Therefore, another sample consisting of MW stars from Gaia DR3 is constructed to map the foreground extinction toward M31.
A total of 368, 388 stars with relative parallax errors less than 0.5 and distances under 50 kpc in the sky area of M31 are selected. 
The observed color indexes from 2MASS, PS1, and Gaia DR3 are used to calculate foreground extinction, ensuring that photometric errors are below 0.05 mag \citep{2019ApJ...877..116W}.


\section{Method} \label{sec:method}

In this work, we map the extinction in M31 by fitting the multiband photometric data of individual stars, a method adopted by \citet{2012ApJ...757..166B}, \citet{2014MNRAS.443.1192C}, \citet{2021ApJ...906...47G}, \citet{2022MNRAS.511.1317C}, etc.

The observed color index between two bands, $\lambda_1$ and $\lambda_2$, can be expressed as: 
\begin{equation}
    (\lambda_1 - \lambda_2) = (\lambda_1 - \lambda_2)_0 + (A_{\lambda_1} - A_{\lambda_2}) = (\lambda_1 - \lambda_2)_0 + A_V(A_{\lambda_1}/A_V - A_{\lambda_2}/A_V),
\end{equation}

where $(\lambda_1 - \lambda_2)_0$ represents the intrinsic color index in bands $\lambda_1$ and $\lambda_2$, while $A_{\lambda_1}$, $A_{\lambda_2}$, and $A_V$ are the absolute extinctions in bands $\lambda_1$, $\lambda_2$, and the $V$ band, respectively. 
The terms $A_{\lambda_1}/A_V$ and $A_{\lambda_2}/A_V$ describe the wavelength dependence of the interstellar extinction, also known as the extinction law or extinction curve, in bands $\lambda_1$ and $\lambda_2$.

\subsection{The Relative Extinction values}\label{Alambda_Av}

The relative extinction values, $A_\lambda/A_V$, for individual bands used in this work are computed based on extinction curves toward different sight lines in M31, as derived by \citet{2022ApJS..259...12W}. 
These extinction curves are modeled using a silicate-graphite dust model with a dust size distribution described by $dn/da \sim a^{-\alpha} \exp(-a/0.25)$, where $a$ is the dust grain radius (assumed to be spherical) and $\alpha$ is the power-law index.
For regions where detailed extinction curves are unavailable, we adopt the average extinction curve derived by \citet{2022ApJS..259...12W}, with a power index $\alpha$ of 3.35, to obtain the relative extinction values, $A_\lambda/A_V$, for each band. 
The calculated $A_\lambda/A_V$ values for the average extinction curve in M31 are listed in Table \ref{tab:extinction_law}.

\begin{deluxetable}{cccc}
    \tablecaption{General extinction predicitons for M31 and MW in multiple bands from UV to IR. \label{tab:extinction_law}}
		\tablehead{	
		\colhead{Band} & \colhead{\hspace{2cm}$\lambda_{\rm eff}$$^a$} & \colhead{\hspace{2cm}$A_{\lambda}/A_V$ (M31)} & \colhead{\hspace{2cm}$A_{\lambda}/A_V$ (MW)} \\
		\colhead{} & \colhead{\hspace{2cm}$(\mu {\rm m})$} & \colhead{\hspace{2cm}\citet{2022ApJS..259...12W}} & \colhead{\hspace{2cm}\citet{2023ApJ...950...86G}, $R_V = 3.1$}
		}
	\startdata	                
	$g$/PS1 & \hspace{2cm}0.481 & \hspace{2cm}1.153 & \hspace{2cm}1.178\\
	$r$/PS1 & \hspace{2cm}0.616 & \hspace{2cm}0.884 & \hspace{2cm}0.864\\
	$i$/PS1 & \hspace{2cm}0.750 & \hspace{2cm}0.704 & \hspace{2cm}0.643\\
	$z$/PS1 & \hspace{2cm}0.867 & \hspace{2cm}0.590 & \hspace{2cm}0.510\\
	$y$/PS1 & \hspace{2cm}0.961 & \hspace{2cm}0.515 & \hspace{2cm}0.422\\
	$J$/UKIRT & \hspace{2cm}1.248 & \hspace{2cm}0.358 & \hspace{2cm}0.265\\
	$H$/UKIRT & \hspace{2cm}1.631 & \hspace{2cm}0.237 & \hspace{2cm}0.169\\
	$K$/UKIRT & \hspace{2cm}2.201 &  \hspace{2cm}0.143 & \hspace{2cm}0.102\\
	$G_{\rm BP}$/GAIA & \hspace{2cm}0.504 & \hspace{2cm}1.098 & \hspace{2cm}1.106\\
	$G_{\rm RP}$/GAIA & \hspace{2cm}0.769 & \hspace{2cm}0.684 & \hspace{2cm}0.619
\enddata
	\tablecomments{$^a$ Effective wavelengths of multiple bands used in this work refer to the SVO Filter Profile Service (http://svo2.cab.inta-csic.es/theory/fps/, \citealt{2012ivoa.rept.1015R}).}
	\end{deluxetable}

\subsection{The intrinsic color indexes}\label{subsec:E0}

The intrinsic color indexes between bands $\lambda_1$ and $\lambda_2$ for tracers in M31 are based on empirical and the theoretical stellar loci in multi-dimensional color space. 
Due to the limited sample of unreddened or slightly reddened stars in M31, we cannot construct the empirical stellar loci using M31 member stars. 
Therefore, we employ two other methods to construct stellar loci for the sample in M31, which are described in Section \ref{subsub:apogee} and Section \ref{subsub:parsec}.

\subsubsection{The intrinsic color indexes based on stars in the MW}\label{subsub:apogee}

Following \citet{2012ApJ...757..166B}, the effective temperature influences colors more significantly than metallicity for most stars. 
Consequently, stars in the MW can be adopted to construct the stellar loci in M31 instead of the member stars.
In this work, we utilize the Apache Point Observatory Galaxy Evolution Experiment (APOGEE, \citealt{2011AJ....142...72E}), a near-infrared stellar spectroscopic survey providing detailed stellar atmospheric parameters, including effective temperature ($T_{\rm eff}$), surface gravity (log $g$), metallicity ([M/H]), and chemical abundances.

By cross-matching stars in the APOGEE catalog with photometric data from the Two Micron All Sky Survey (2MASS, \citealt{2003yCat.2246....0C}), the PS1 Survey \citep{2016arXiv161205560C}, and Gaia DR3, the color indexes $g-r$, $r-i$, $i-z$, $z-y$, $y-J$, $J-H$, $H-K_{\rm s}$, and $G_{\rm BP}-G_{\rm RP}$ for the reference sample are obtained.
Since 2MASS and UKIRT filter systems differ \citep{2009MNRAS.394..675H,2021ApJ...923..232R}, it is necessary to convert 2MASS magnitudes to the UKIRT system with equations \citep{2009MNRAS.394..675H}:
\begin{equation}
	J_{\rm W} = J_2 - 0.065(J_2 - H_2),
	\label{eq:J2JW}
\end{equation}
\begin{equation}
	H_{\rm W} = H_2 + 0.070(J_2 - H_2) - 0.030,
	\label{eq:H2HW}
\end{equation}
\begin{equation}
	K_{\rm W} = K_2 + 0.010(J_2 - K_2),
	\label{eq:K2KW}
\end{equation}
where $J_{\rm W}$, $H_{\rm W}$, and $K_{\rm W}$ are the photometric magnitudes in the filter system of the UKIRT/WFCAM, while $J_2$, $H_2$, and $K_2$ are those of the 2MASS.
In line with \citet{2022MNRAS.511.1317C}, we restrict the reference sample to stars with photometric errors smaller than 0.05 mag in all bands and low line-of-sight extinction from the extinction map of \citet{1998ApJ...500..525S}.

\begin{figure}[ht!]
    \centering
	\includegraphics[scale = 0.5]{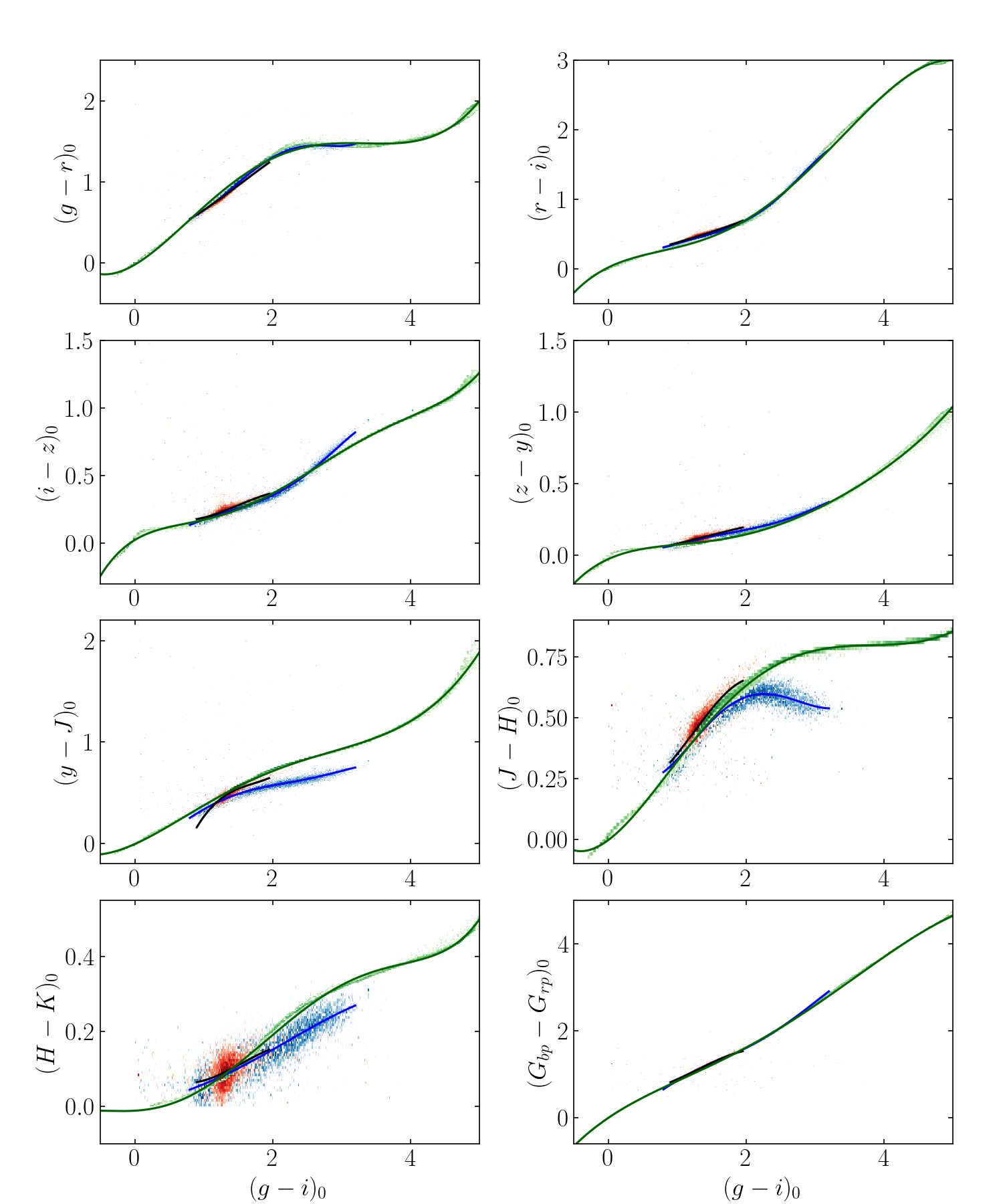}	
	\caption{Stellar number density distributions in individual color-color diagrams for dwarfs (blue contours) and giants/supergiants (red contours) from the APOGEE catalog, along with theoretical distributions generated using the \emph{CMD 3.7} web interface (green contours).
	The blue, black and green lines represent fifth-order, third-order, and fifth-order polynomial fits, respectively, to the median values obtained after 3-$\sigma$ clipping.\label{fig:stellar_loci}}
\end{figure}

Although effective temperature primarily affects intrinsic colors, the brightness in the $H$ band differs between dwarfs and giants due to surface gravity differences\footnote{Higher surface gravity makes the collision rate between atoms higher and thus molecules form more easily, which leads to absorption in the $H$ band for dwarfs \citep{1988PASP..100.1134B}.}, as argued by \citet{1988PASP..100.1134B} and \citet{2021ApJ...907...18R}. 
We therefore construct separate stellar loci for dwarfs and giants/supergiants, shown in Figure \ref{fig:stellar_loci}.
It is assumed that the intrinsic color index $(g-i)_0$ is the independent variable.
The blue contours in Figure \ref{fig:stellar_loci} represent the stellar number density distributions in the individual color-color planes for dwarfs in the APOGEE reference sample.
The median values, obtained after 3-$\sigma$ clipping, are fitted with fifth-order polynomials, as described in \citet{2021ApJ...906...47G} and \citet{2022MNRAS.511.1317C}.
Giants and supergiants are shown with red contours, and the stellar loci are derived with the third polynomial fits \citep{2021ApJ...906...47G}.
Table \ref{tab:fit_parameters} lists the corresponding fit parameters and residuals, which indicates that the polynomial fits effectively capture the underlying trends in the data.

\begin{deluxetable}{cccccccccc}
    \tablecaption{Coefficients of polynomial fits to the stellar loci based on the reference stars in the MW$^a$ and the theoretical stellar loci based on the stellar evolutionary model$^b$. \label{tab:fit_parameters}}
		\tablehead{	
		\colhead{Band} & \colhead{$g-r$} & \colhead{$r-i$} & \colhead{$i-z$} & \colhead{$z-y$} & \colhead{$y-J$} & \colhead{$J-H$} & \colhead{$H-K$} & \colhead{$G_{\rm BP}-G_{\rm RP}$} 
		}
	\startdata	                
	D$_5$ & 0.0428 & -0.0499 & -0.0154 & -0.0136 & -0.0162 & -0.0079 & -0.0000 & -0.013 \\
	D$_4$ & -0.3098 & 0.4158 & 0.1218 & 0.1371 & 0.1565 & 0.1331 & -0.0011 & 0.0659 \\
	D$_3$ & 0.6345 & -1.2117 & -0.3102 & -0.5098 & -0.5242 & -0.7300 & 0.0029 & 0.1001 \\
	D$_2$ & -0.1690 & 1.6208 & 0.2977 & 0.8766 & 0.6469 & 1.6149 & 0.0174 & -0.9315 \\
	D$_1$ & 0.0387 & -0.7356 & 0.1064 & -0.5942 & 0.1212 & -1.2311 & 0.0372 & 2.2634 \\
	D$_0$ & 0.4114 & 0.3272 & -0.0269 & 0.1782 & -0.0491 & 0.5497 & 0.0027 & -0.6355 \\
	Maximum Residual & 0.0094 & 0.0195 & 0.0094 & 0.0042 & 0.0033 & 0.0074 & 0.0062 & 0.0336 \\
	Median Residual & 0.0055 & 0.0052 & 0.0032 & 0.0008 & 0.0010 & 0.0026 & 0.0030 & 0.0077 \\
	\hline
	G$_3$ & -0.0835 & 0.0223 & -0.1109 & -0.0141 & 0.4520 & -0.1585 & -0.0520 & -0.2928 \\
	G$_2$ & 0.3940 & -0.0451 & 0.5000 & 0.0597 & -2.3440 & 0.5847 & 0.2429 & 1.1182 \\
	G$_1$ & 0.0190 & 0.3110 & -0.5406 & 0.0384 & 4.2638 & -0.3388 & -0.2784 & -0.6417 \\
	G$_0$ & 0.3230 & 0.0922 & 0.3401 & -0.0040 & -2.1093 & 0.2641 & 0.1563 & 0.7005 \\
	Maximum Residual & 0.0248 & 0.0428 & 0.0270 & 0.0096 & 0.1907 & 0.1155 & 0.0619 & 0.2235 \\
	Median Residual & 0.0113 & 0.0102 & 0.0078 & 0.0026 & 0.0103 & 0.0086 & 0.0063 & 0.0272 \\
	\hline
	P$_5$ & -0.0011 & 0.0006 & 0.0042 & 0.0012 & -0.0005 & -0.0008 & 0.0013 & 0.0007 \\
	P$_4$ & 0.0342 & -0.0283 & -0.0515 & -0.0173 & 0.0152 & 0.0159 & -0.0117 & -0.0210 \\
	P$_3$ & -0.2279 & 0.2046 & 0.2161 & 0.0948 & -0.0911 & -0.1019 & 0.0226 & 0.1539 \\
	P$_2$ & 0.4068 & -0.3725 & -0.3281 & -0.1962 & 0.1530 & 0.1990 & 0.0363 & -0.3630 \\
	P$_1$ & 0.4972 & 0.4853 & 0.3087 & 0.2182 & 0.3040 & 0.2116 & 0.0111 & 1.0747 \\
	P$_0$ & -0.0192 & 0.0215 & 0.0231 & -0.0276 & -0.0060 & -0.0025 & -0.0117 & -0.0027 \\
	Maximum Residual & 0.0566 & 0.0589 & 0.0300 & 0.0407 & 0.0661 & 0.0275 & 0.0223 & 0.0352 \\
	Median Residual & 0.0237 & 0.0257 & 0.0054 & 0.0085 & 0.0068 & 0.0048 & 0.0060 & 0.0127
\enddata
	\tablecomments{$^a$ The stellar loci for dwarfs can be expressed as $(\lambda_1-\lambda_2)_0 = {\rm D}_5(g-i)_0^5 + {\rm D}_4(g-i)_0^4 + {\rm D}_3(g-i)_0^3 + {\rm D}_2(g-i)_0^2 + {\rm D}_1(g-i) + {\rm D}_0$, while those for giants/supergiants are  $(\lambda_1-\lambda_2)_0 = {\rm G}_3(g-i)_0^3 + {\rm G}_2(g-i)_0^2 + {\rm G}_1(g-i) + {\rm G}_0$.\\
	$^b$ The theoretical stellar loci  can be expressed as $(\lambda_1-\lambda_2)_0 = {\rm P}_5(g-i)_0^5 + {\rm P}_4(g-i)_0^4 + {\rm P}_3(g-i)_0^3 + {\rm P}_2(g-i)_0^2 + {\rm P}_1(g-i) + {\rm P}_0$.}
	\end{deluxetable}

\subsubsection{The intrinsic color indexes based on the theoretical stellar loci}\label{subsub:parsec}

As shown in Figure \ref{fig:stellar_loci}, the stellar loci for giants and supergiants are limited to a narrow range ($\sim$ $1 < (g-i)_0 < 2$).
However, the sample from \citet{2021ApJ...907...18R} mostly contains evolved stars, rather than main-sequence stars. 
Thus, the stellar loci derived from MW reference stars can only be applied to a limited number of tracers in M31.
To extend this range, we employ stellar evolutionary models to construct theoretical stellar loci.

The theoretical intrinsic magnitudes in the PS1, 2MASS\footnote{The magnitudes from the 2MASS are also converted to the UKIRT system with Equations \ref{eq:J2JW}-\ref{eq:K2KW}.} and Gaia bands are generated from the \emph{CMD 3.7} web interface.
The magnitudes are based on PARSEC and COLIBRI evolutionary tracks, which account for factors such as circumstellar dust, interstellar extinction, long period variability, initial mass function (IMF), age, and metallicity.
The input parameters in this work follow those in Table 1 of \citet{2024ApJ...966...25R}, with slight adjustments. 
Given the uncertain detailed information of the abundance of M31, we adopt [M/H] = 0, as  solar abundances have historically been taken to represent the total interstellar abundances \citep{2005ApJ...622..965L}.
The obtained theoretical stellar loci are also shown in Figure \ref{fig:stellar_loci}, and the corresponding fit parameters are listed in Table \ref{tab:fit_parameters}.

~\\
In this work, we use stellar loci derived from the APOGEE sample in the range $1.3 < (g-i)_0 < 2$ and theoretical stellar loci from the \emph{CMD 3.7} web interface in the range $2 \le (g-i)_0 < 5$ to calculate the extinction in the $V$ band for individual stars in M31.

\subsection{The Construction of the Extinction Map}\label{method_map}

The observed color index can be parameterized by the intrinsic color index $(g-i)_0$ and the extinction value in the $V$ band, $A_V$.

For the 294, 065 tracers with more than two observed color indexes (as mentioned in Section \ref{sec:data}), the best-fitting intrinsic color $(g-i)_0$ and extinction value $A_V$ are determined by minimizing the $\chi^2$ function, defined as: 
\begin{equation} 
	\chi^2 = \frac{1}{N_{\rm data} - N_{\rm para}} \sum\limits_{i=1}^{N_{\rm data}} \left[ \frac{(\lambda_{1} - \lambda_{2})_{\rm obs(i)} - (\lambda_{1} - \lambda_{2})_{\rm mod(i)}}{\sigma_i} \right]^2,
\end{equation} where $(\lambda_{1} - \lambda_{2})_{\rm obs(i)}$ is the observed color index (e.g., $g-r$, $r-i$, $i-z$, etc.), $(\lambda_1 - \lambda_2)_{\rm mod(i)}$ is the modeled color index, and $\sigma_i$ is the associated uncertainty.

\begin{figure}[ht!]
    \centering
	\includegraphics[scale = 0.8]{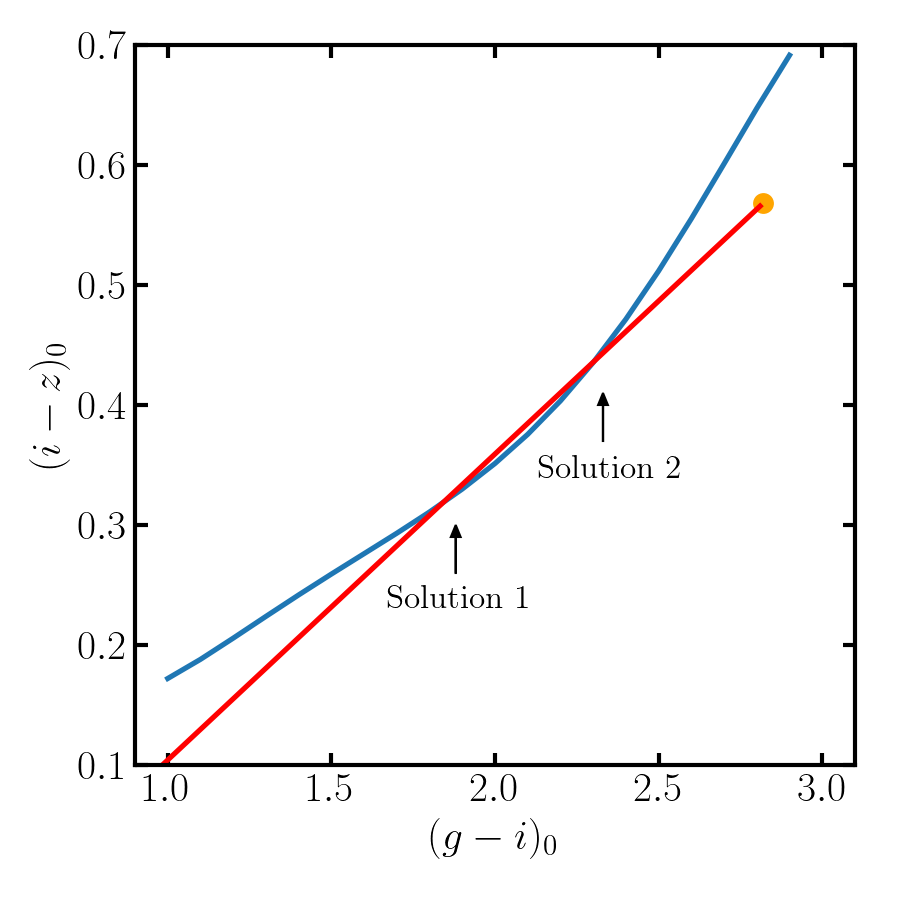}	
	\caption{Example of a multi-solution situation. 
	The orange point represents the observed colors of a tracer in color-color space. 
	The red line is parallel to the reddening vector and intersects the stellar locus at two points (Solution 1 and Solution 2).
	 Solution 1 corresponds to a star at a shorter distance with greater extinction, while Solution 2 represents a more distant star with smaller extinction.
	\label{fig:method_OnePoint} }
\end{figure}

For the 3430 tracers where only the color index $(g-i)$ and one other observed color index are available, the location of the tracer in color-color space is shifted along the reddening vector, and the intersection with the stellar locus indicates the intrinsic color. 
The extinction in the $V$ band is then calculated by comparing the derived intrinsic color with the observed one.
However, due to different possible combinations of distance and extinction, the observed tracer may correspond to multiple intrinsic stellar types, resulting in more than one intersection with the stellar loci, as shown in Figure \ref{fig:method_OnePoint}.
In such cases, we choose the solution with smaller extinction, which aligns with the external galaxy distance.
Considering that the reliability of the results depends on the number of the observed photometry, the $A_V$ values obtained in this situation are assigned a weight of 0.6 compared to those from the tracers with more than two observed color indexes.

For the 208, 723 tracers with two observed color indexes, i.e., $(\lambda_1 - \lambda_2)_{\rm obs(1)}$ and $(\lambda_1 - \lambda_2)_{\rm obs(2)}$, $(g-i)_0$ and $A_V$ are derived by solving the following system of equations: 
\begin{equation}
    \begin{cases}
    (\lambda_1 - \lambda_2)_{\rm obs(1)} =  (\lambda_1 - \lambda_2)_{\rm int(1)}[(g-i)_0] + A_V(A_{\lambda_1}/A_V - A_{\lambda_2}/A_V)_{(1)}, \\
    (\lambda_1 - \lambda_2)_{\rm obs(1)} =  (\lambda_1 - \lambda_2)_{\rm int(2)}[(g-i)_0] + A_V(A_{\lambda_1}/A_V - A_{\lambda_2}/A_V)_{(2)}, 
	\end{cases}
\end{equation}
where $(\lambda_1 - \lambda_2)_{\rm int}$ is a cubic or fifth-order function of $(g-i)_0$, and $A_\lambda/A_V$ refers to the extinction law in M31 (as discussed in Section \ref{Alambda_Av}). 
We retain the real solution(s) of $(g-i)_0$ that fall within the range of $1.3 < (g-i)_0 < 5$.
If multiple solutions exist, we adopt the one with the smaller $A_V$ value.
The derived results in this case are given a weight of 0.8.

For the 316, 971 tracers with less than two observed color indexes, we do not calculate the extinction values.

Once the extinction values $A_V$ for individual tracers are computed, the extinction distribution in M31 can be constructed.
The sky area is divided into subfields (pixels) with a resolution of approximately 50 arcsec.
When calculating the extinction for each pixel, the distribution of stars along the sight line must be considered.
According to \citet{2010A&A...509A..91T} and \citet{2011MNRAS.413.1548C}, the dust disk in M31 is significantly thinner than the stellar disks\footnote{The scaleheights for the thin and thick stellar disks in M31 are $1.1 \pm 0.2$ and $2.8 \pm 0.6$ kpc, respectively \citep{2011MNRAS.413.1548C}, while the scaleheight of the dust disk in M31 is set to 0.24 kpc \citep{2010A&A...509A..91T}.}.
As described in \citet{2015ApJ...814....3D}, the thickness of the dust disk in M31 can be considered negligible.
Consequently, tracers with significant extinction can be assumed to lie behind the dust layer. 
However, for tracers with little or negligible extinction (i.e., $A_V < 0.1$ mag), it is essential to account for the relative positions with respect to the dust.
Referring to the discussion of the fraction of stars behind the dust layer ($f_{\rm red}$) in \citet{2015ApJ...814....3D}, it can be assumed that approximately half of these low-extinction sources are located in front of the dust (and thus remain unaffected), while the other half are behind the dust (and experience only minimal extinction).
Based on the previous discussion, the stellar count in each pixel should be the actual number of stars, subtracting half of the number of low-extinction stars.
For each pixel, tracers with extinction values exceeding the 3$\sigma$ limit are excluded. 
The weighted average $A_V$ value of the remaining tracers, along with its variance, represents the extinction value and uncertainty for that pixel.

\section{Results and discussions} \label{sec:re}

\subsection{The MW foreground extinction distribution toward M31}\label{subsec:foreground}

Before constructing the extinction distribution of external galaxies, the foreground extinction from the MW must be considered. 
Therefore, we first map the MW extinction in the M31 sky area to remove the contribution from the total extinction in M31.

The foreground extinction distribution is created using a sample of MW stars from Gaia DR3, as detailed in Section \ref{sec:data}. 
With the combination of the stellar loci derived in Section \ref{subsub:apogee} and the average extinction law for diffuse regions in the MW ($R_V = 3.1$, \citealt{2023ApJ...950...86G}, see Table \ref{tab:extinction_law} for details), we fit the observed color indexes of individual MW stars. 
The foreground extinction distribution is constructed using the same methodology described in Section \ref{sec:method}.

\begin{figure}[ht!]
    \centering
	\includegraphics[scale = 0.8]{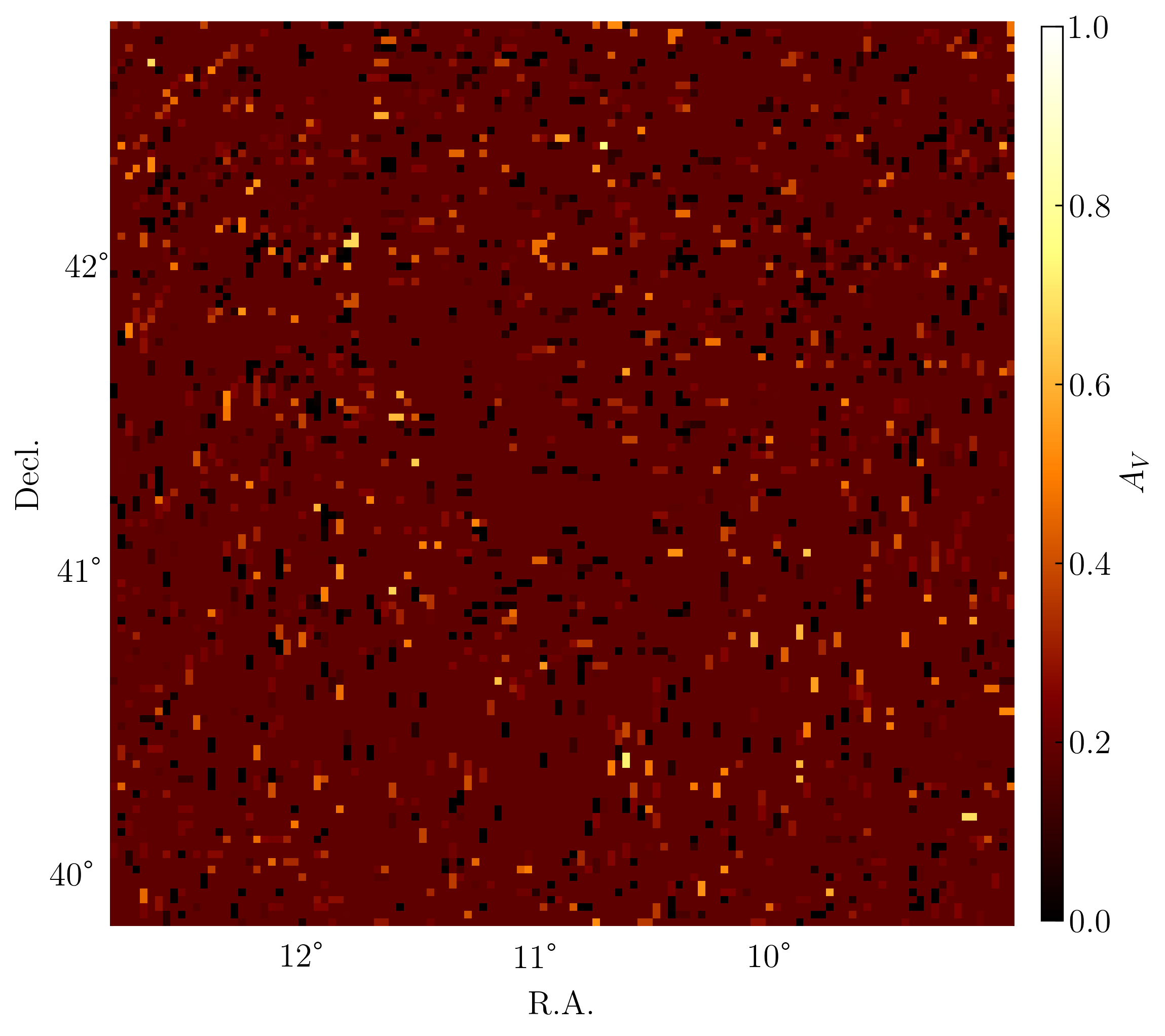}	
	\caption{Foreground extinction distribution toward M31 with a resolution of approximately 1.7 arcmin.
	 \label{fig:foreground_extinction}}
\end{figure}

We analyzed approximately 0.32 million MW stars toward M31. 
Figure \ref{fig:foreground_extinction} shows the foreground extinction distribution toward M31 with a resolution of approximately 1.7 arcmin. 
There are no obvious structural features visible in the figure, suggesting a random distribution of foreground extinction.
The average foreground extinction derived in this work is $A_V \approx 0.185$ mag, which corresponds to the uniform foreground extinction $E(B-V) \approx 0.06$ mag, assuming an $R_V = 3.1$ extinction curve \citep{1998ApJ...500..525S,2011ApJ...737..103S,2012AJ....144..142B,2020ApJ...905L..20R}, used in previous studies (e.g., \citealt{2015ApJ...815...14C,2022ApJS..259...12W}).

\subsection{Extinction distribution in M31}

\begin{figure*}[ht]
	\centering
	\begin{minipage}{0.45\textwidth}
		\centering
		\raisebox{-0cm}{\includegraphics[height=7.5cm]{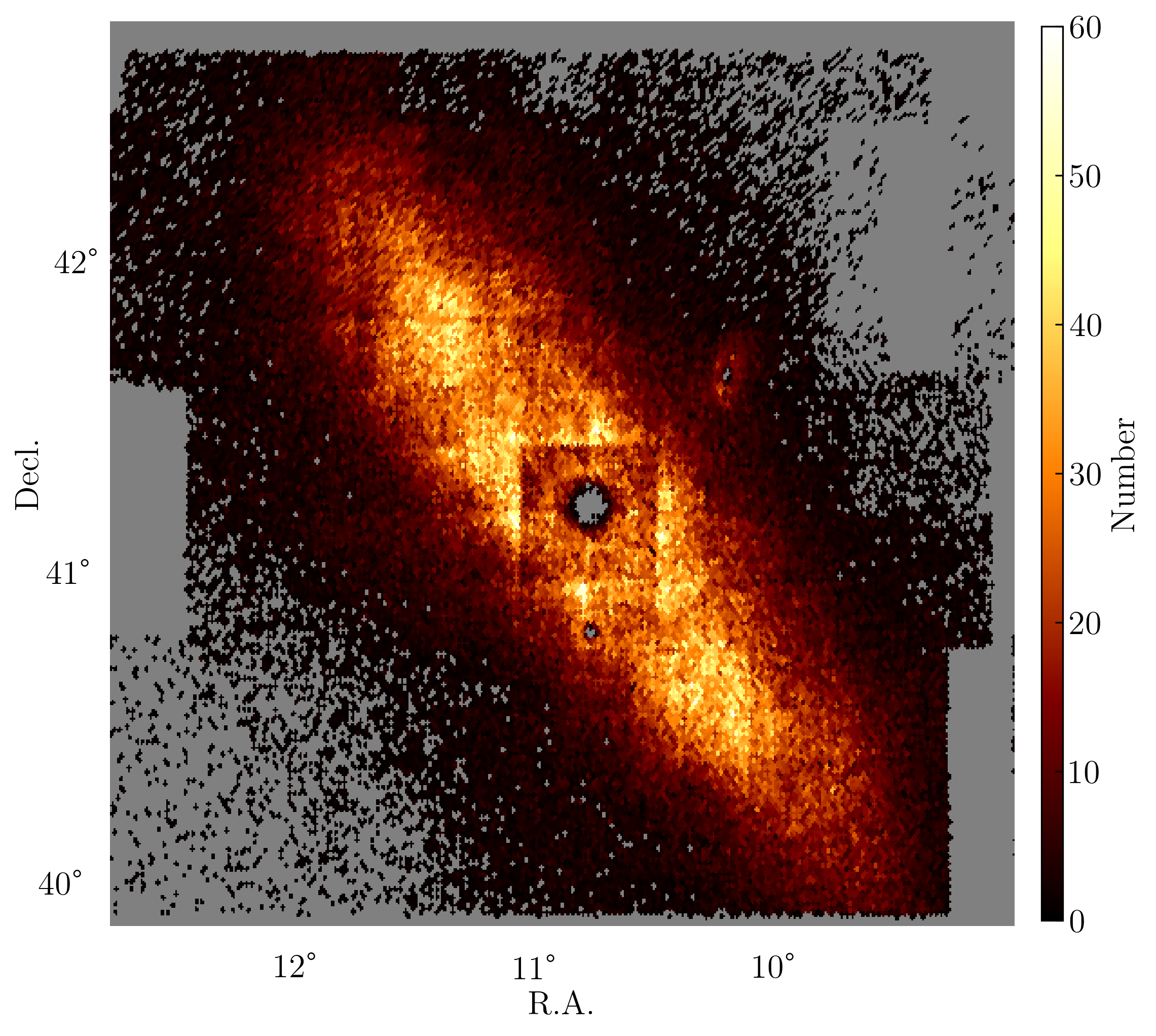}}
	\end{minipage}\hspace{0.01\textwidth}
	\begin{minipage}{0.45\textwidth}
		\centering
		\includegraphics[height=7.5cm]{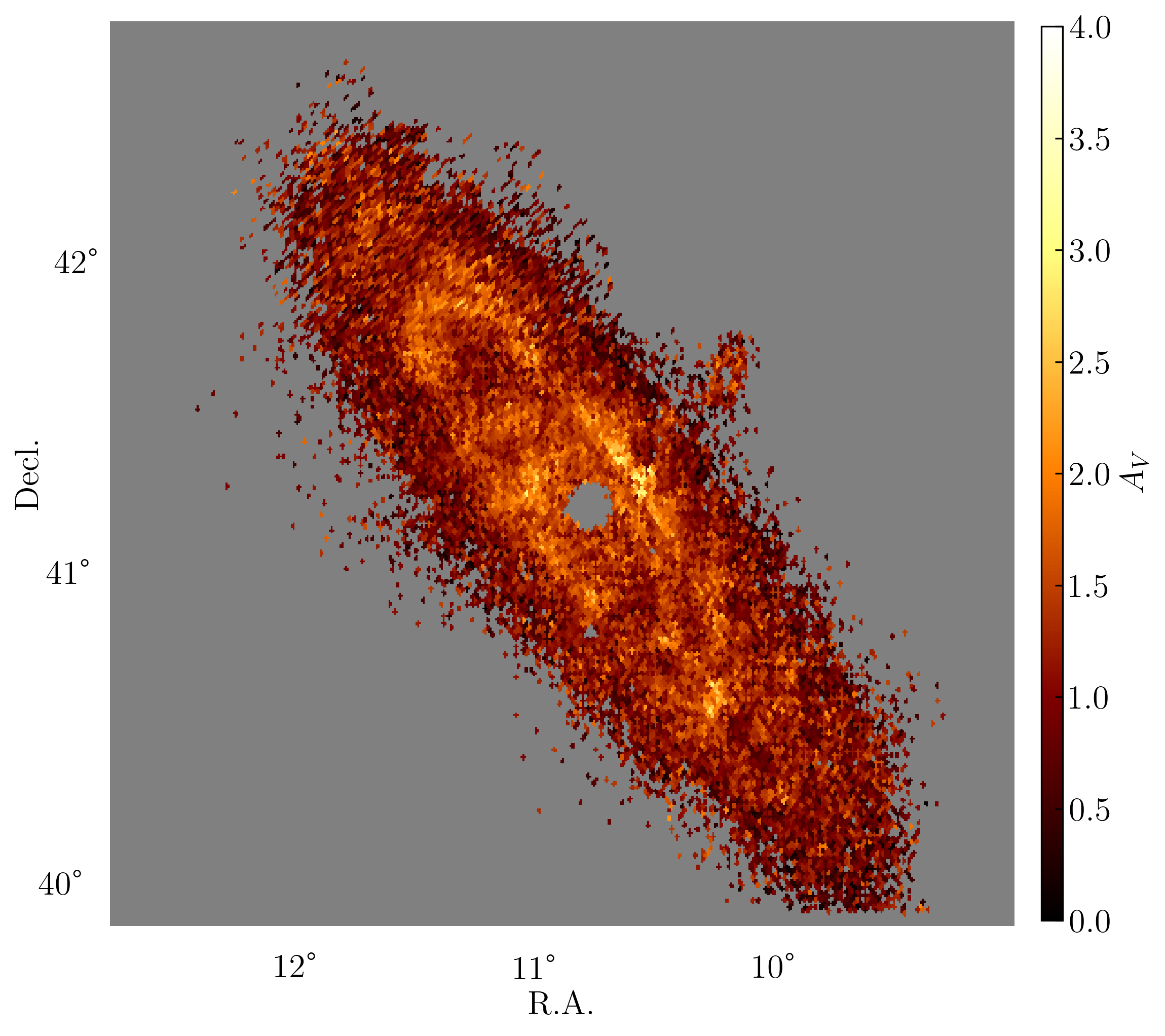}
	\end{minipage}
	\caption{Left panel: Distribution of the stellar number in each pixel.
	Right panel: Extinction distribution in M31 with a resolution of approximately 50 arcsec.
	\label{fig:M31_extinction}}
\end{figure*}

With the foreground extinction map derived in Section \ref{subsec:foreground}, a final sample of approximately 0.34 million stars is obtained.
The left panel of Figure \ref{fig:M31_extinction} shows the stellar distribution, where each pixel contains up to 60 stars, with a median of 3.5 stars and an average of 8.32 stars per pixel.
The Kolmogorov-Smirnov (KS) test is performed, and it is found that setting the minimum number of stars per pixel to 8 or more does not significantly affect the overall extinction distribution of M31.
Therefore, we impose a requirement of at least 8 stars per pixel when constructing the extinction distribution, as shown in the right panel of Figure \ref{fig:M31_extinction}.

The map has a resolution of approximately 50 arcsec ($\sim 189$ pc) and covers around 10 deg$^2$ of the sky toward M31. 
Notably, the extinction in the disk region of M31 is significantly higher than in the surrounding areas, with distinct spiral features visible. 
The spiral arms, especially those located at the northeastern and southwestern parts of the galaxy, are well aligned with known star-forming regions in M31.
Several spiral arms can be identified at approximately 2.8 kpc, 5.6 kpc, 11.2 kpc, and 15.6 kpc from the galactic center, corresponding well with the positions of spiral arms observed in the optical band, confirming the reliability of the extinction map.
In contrast, the central region shows a clear gap in extinction, which is due to a lack of observable sources in this area.
The outer regions of the map show increased noise, which arises from random fluctuations caused by the limited number of observed sources in these areas.
In the densest parts of the spiral arms, extinction values reach up to $A_V \approx 3-4$ mag, while in the inter-arm regions and outer disk, the extinction decreases to below 0.5 mag. 
This radial and structural variation in extinction closely matches the distribution of dust and star-forming regions across M31, reinforcing the accuracy and depth of the map.
The average extinction value in the $V$ band is $A_V = 1.17$ mag, which is consistent with previous studies \citep{2022ApJS..259...12W}.

The extinction map generated in this work as well as the foreground extinction map described in Section \ref{subsec:foreground} are publicly accessible in Zenodo (\citealt{wang_2025_14887803}, \url{https://doi.org/10.5281/zenodo.14887803}).
Additionally, the Zenodo repository provides a simple Python procedure that returns extinction values in the $V$ band for given input positions (RA and Dec.). 
An example of how to use this procedure is also included.


\subsection{Comparison with other works}

\begin{figure*}[ht]
	\centering
	\begin{minipage}{0.45\textwidth}
		\centering
		\raisebox{-0.6cm}{\includegraphics[height=7.5cm]{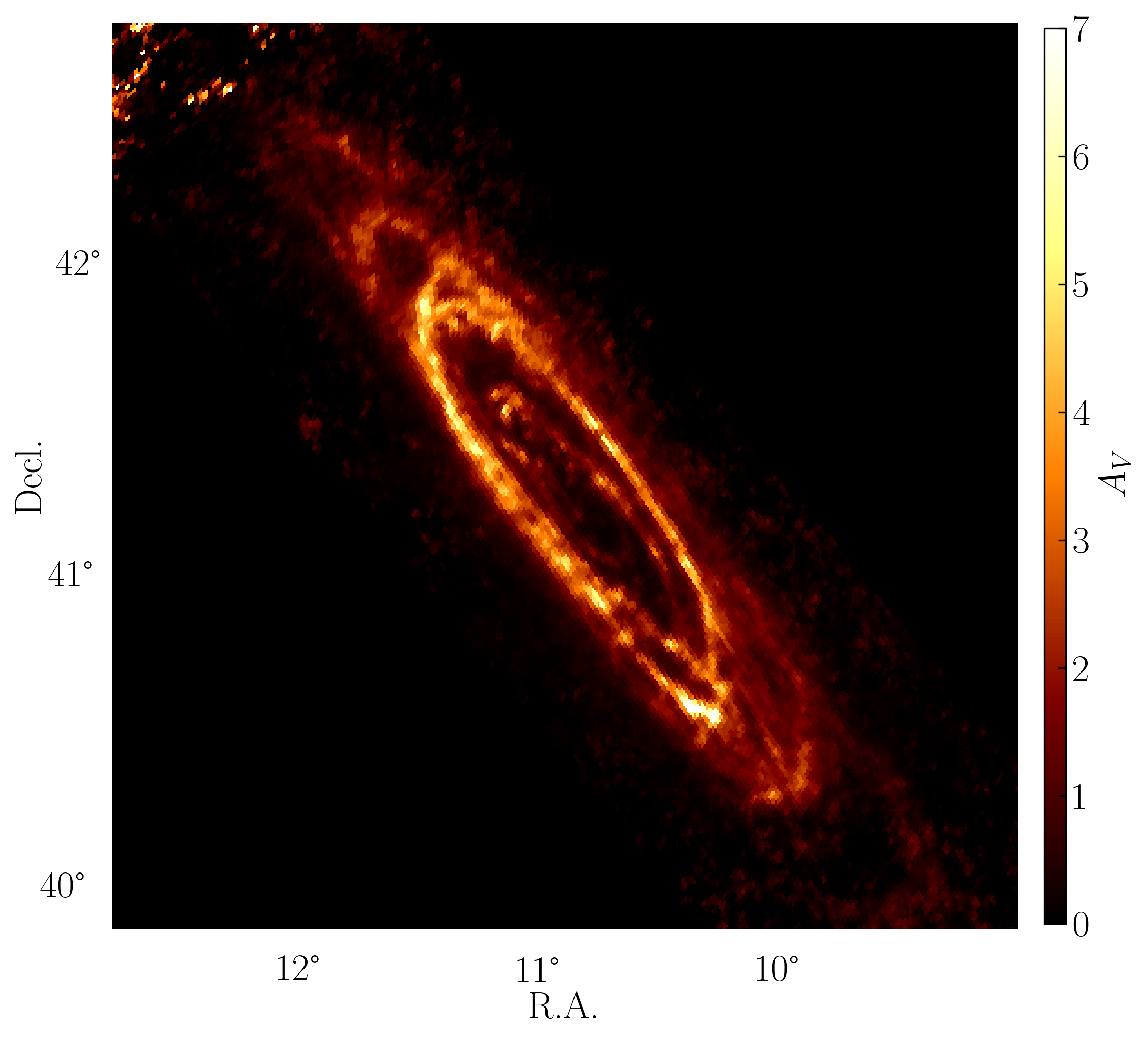}}
		\label{fig:figure1}
	\end{minipage}\hspace{0.01\textwidth}
	\begin{minipage}{0.45\textwidth}
		\centering
		\includegraphics[height=7.5cm]{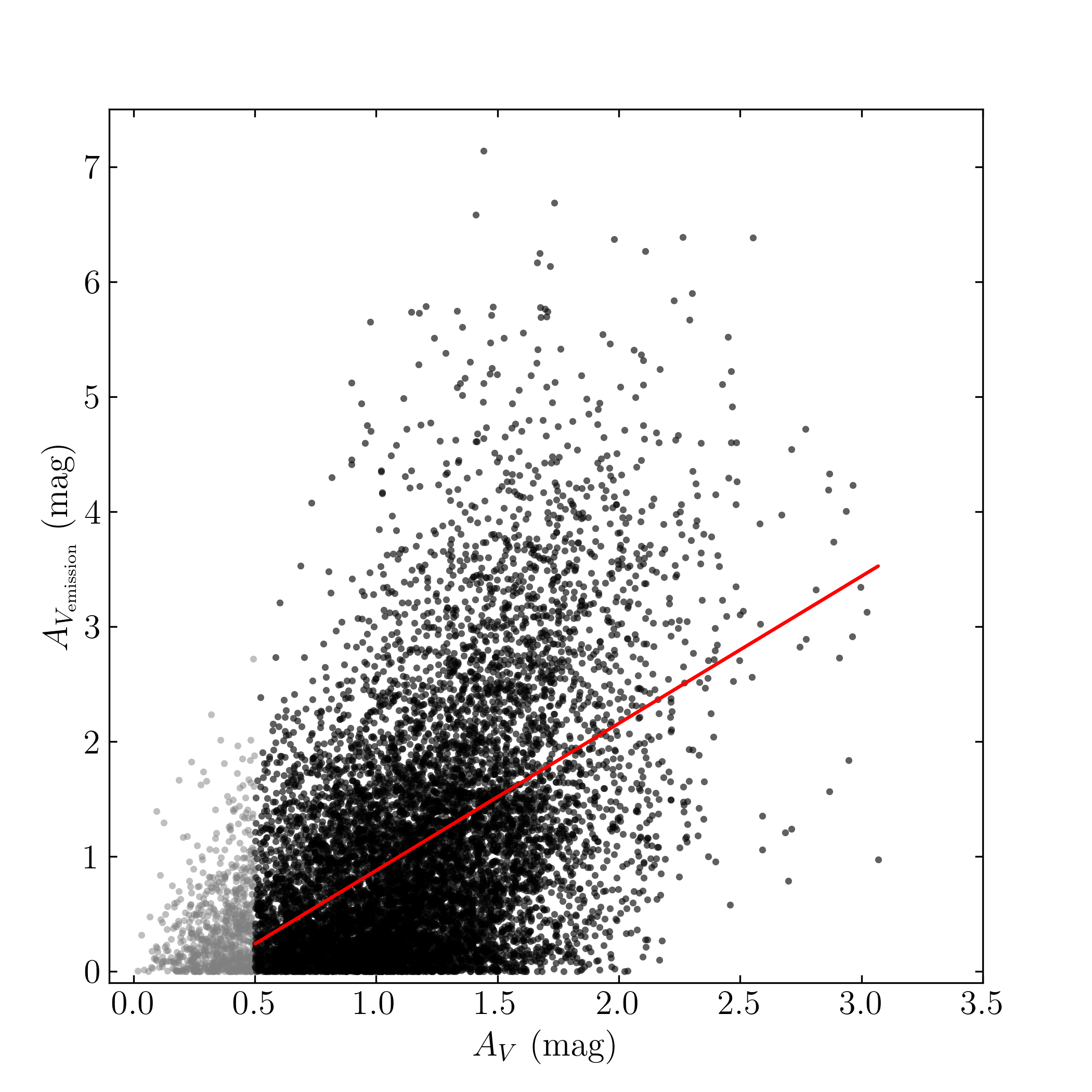}
		\label{fig:figure2}
	\end{minipage}
	\caption{Comparison between the extinction distributions derived in this work and that inferred from \citet{2014ApJ...780..172D}. 
	Left panel: Extinction distribution calculated from the dust mass surface density map based on infrared emission \citep{2014ApJ...780..172D}, with the same resolution as Figure \ref{fig:M31_extinction}.
	Right panel: Pixel-by-pixel comparison between $A_V$ values derived in this work (x-axis) and the resolution-matched $A_{V_{\rm emission}}$ values from the emission-based map (y-axis). 
	The red line shows the linear regression fit for$A_V > 0.5$ mag, given by the equation $A_{V_{\rm emission}} = 1.28A_V - 0.40$.\label{fig:compare}}
\end{figure*}

As stated in Section \ref{sec:intro}, extinction distributions can also be determined through infrared emission based on physical dust models. 
By converting the dust surface density ($\Sigma_{M{\rm d}}$) map from \citet{2014ApJ...780..172D} to extinction in the $V$ band with the relation $A_{V_{\rm emission}} = 0.74(\frac{\Sigma_{ M \rm d}}{10^5M_{\odot}{\rm kpc}^{-2}})$ \citep{2007ApJ...657..810D}, an emission-based extinction distribution in M31 is inferred, shown in the left panel of Figure \ref{fig:compare} with the same resolution as the extinction map shown in Figure \ref{fig:M31_extinction}. 
Despite the independent nature of these methodologies, notable morphological agreement appears between the two extinction maps, especially in the spiral regions of M31.

A quantitative assessment is conducted through a pixel-by-pixel comparison between the extinction derived in this work ($A_V$) and the emission-based extinction from \citet{2014ApJ...780..172D} ($A_{V_{\rm emission}}$), as displayed in the right panel of Figure \ref{fig:compare}.
As noted in \citet{2015ApJ...814....3D}, the method adopted in this work is limited by the accuracy of the stellar loci, which can lead to larger errors in low-extinction regions.
As a result, the comparison is restricted to points with $A_V > 0.5$ mag as described in \citet{2015ApJ...814....3D}.
A fit of the data yields the relation $A_{V_{\rm emission}} = 1.28A_V - 0.40$, represented by the red line in the right panel of Figure \ref{fig:compare}.

As shown by the relationship described above, extinction values derived from emission are generally higher than those obtained in this work. 
This discrepancy may arise due to several factors.
On one hand, as discussed by \citet{2015ApJ...814....3D}, wide-field background subtraction and calibration uncertainties can introduce systematic offsets, particularly in regions with low extinction. 
Additionally, unresolved dust components with varying temperatures can complicate infrared emission interpretations (e.g., \citealt{2011A&A...536A..88G}), making emission-based extinction maps less reliable in low-extinction areas.
On the other hand, differing assumptions about dust distribution may also contribute to the observed differences in extinction values.
The dust size distribution adopted by \citet{2014ApJ...780..172D} corresponds to the diffuse interstellar medium in the solar neighborhood \citep{2001ApJ...548..296W}, which is commonly used in modeling extinction curves and infrared emission. 
In contrast, the extinction law applied in this work is based on a silicate-graphite dust model with a size distribution of $dn/da \sim a^{-\alpha}$exp($-a/0.25$).
Moreover, while \citet{2014ApJ...780..172D} assumed a uniform dust distribution for M31, this work employs extinction curves based on distinct dust size distributions. 
Furthermore, while \citet{2014ApJ...780..172D} reports extinction values as high as $A_V \approx 7$ mag, the densest regions in this study show extinction values of up to 3-4 mag. 
This difference arises because the extinction values in our study are based on direct observations, and sources in regions with higher extinction are not detectable in our data.

It is worth noting that \citet{2015ApJ...814....3D} also conducted a study on the extinction distribution in M31 and found that the extinction values reported by \citet{2014ApJ...780..172D} were approximately 2.5 times larger than those measured in their work.
Given that the extinction values derived in this work are generally higher than those obtained by \citet{2014ApJ...780..172D}, by a factor of approximately 1.28, it suggests that our results are about half of those reported by \citet{2015ApJ...814....3D}.
This discrepancy may be attributed to several factors. 
First, the work of \citet{2015ApJ...814....3D} focused on a more localized portion of M31 (approximately one-third of M31's star forming disk), while this work covers the entire galaxy. 
Additionally, variations in dust distribution, observational techniques, and assumptions regarding the dust model could explain these differences.
However, due to access limitations, we are unable to obtain the relevant data from \citet{2015ApJ...814....3D} for a direct comparison.
Once their data become available, a more thorough comparison could help clarify the reasons for this discrepancy and provide deeper insights into the extinction properties of M31.

The extinction distribution based on stellar observations in this work serves as a useful tool for correcting extinction in future observations (e.g., CSST, \citealt{2021CSB...111.111C}) and for further studies of M31's stellar population. 
Increased sample size and broader wavelength coverage are expected to enhance the resolution and accuracy of the extinction map. 
Additionally, extinction distributions from stellar observations play a critical role in refining dust models for M31.


\section{Conclusion} \label{sec:conclusion}

The member stars identified by \citet{2021ApJ...907...18R} are chosen as extinction tracers in M31.
The sample covers almost the entire sky area of M31, excluding the bulge region. 
By utilizing photometric data from UKIRT/WFCAM, PS1, and Gaia DR3, an extinction distribution in M31 is constructed, representing the largest coverage area among stellar-based extinction maps for M31. 
The main results are summarized as follows:

1. A foreground extinction distribution toward M31 with a resolution of approximately 1.7 arcmin is successfully constructed with a sample of MW stars in the M31 sky region from Gaia DR3. 
The derived average foreground extinction is $A_V \approx 0.185$ mag, consistent with the uniform foreground extinction values applied in previous studies. 
This foreground extinction map serves as a crucial correction factor in constructing the extinction distribution for M31.

2. The extinction distribution in M31 derived in this work covers a sky area of approximately 10 deg$^2$ with a resolution of $\sim$ 50 arcsec.
The map captures key morphological features of M31, such as its prominent ring structures, providing an accurate depiction of dust distribution. 
The average extinction, $A_V \approx$ 1.17 mag, is consistent with previous studies.



\acknowledgments 
We sincerely appreciate the valuable and constructive report provided by the anonymous reviewer.
It is a pleasure to thank Profs. Biwei Jiang and Haibo Yuan for their invaluable discussions and insights. 
This work is supported by the National Natural Science Foundation of China under project 12303030 and the Shandong Provincial Natural Science Foundation under project ZR2024QA236.
Additional support is provided by the National Natural Science Foundation of China (12203025, 12133002, U2031209, 12173034, 12322304), the Shandong Provincial Natural Science Foundation (ZR2022QA64), the CSST Project CMS-CSST-2021-A09, and Shandong Provincial University Youth Innovation and Technology Support Program (No. 2022KJ138).
The numerical computations were conducted on the Qilu Normal University High Performance Computing, Jinan Key Laboratory of Astronomical Data.
This work is dedicated to the Department of Astronomy at Beijing Normal University, with our deepest gratitude and lasting remembrance.

\bibliography{paper}{}
\bibliographystyle{aasjournal}



\end{CJK*}
\end{document}